\documentstyle[aps,epsfig,multicol]{revtex}

\newcommand{\vecs}[1]{{\bf #1}}

\newcommand{\be}{\begin{equation}}
\newcommand{\ee}{\end{equation}}
\newcommand{\bea}{\begin{eqnarray}}
\newcommand{\eea}{\end{eqnarray}}
\newcommand{\lb}{\left[}
\newcommand{\rb}{\right]}
\newcommand{\lp}{\left(}
\newcommand{\rp}{\right)}

\def\breakon{\end{multicols}\widetext\vspace{-.2cm}
\noindent\rule{.48\linewidth}{.3mm}\rule{.3mm}{.3cm}\vspace{.0cm}}
 
\def\breakoff{\vspace{-.2cm}
\noindent
\rule{.52\linewidth}{.0mm}\rule[-.27cm]{.3mm}{.3cm}\rule{.48\linewidth}{.3mm}
\vspace{-.3cm}
\begin{multicols}{2}
\narrowtext}
 
\begin{document}
\draft
 
\title{Internal waves and synchronized precession in a cold vapor}
\author{M. \"O. Oktel$^a$ and L. S. Levitov$^b$}
\address{$^a$ Department of Physics, Ohio State University, 174 West 18 Avenue, Columbus, OH 43210\\
$^b$ Department of Physics, 
Massachusetts Institute of Technology, 77 Massachusetts Ave, Cambridge, MA 02139}

\date{\today}
 
\maketitle
 
\begin{abstract}  
    Exchange in a Boltzmann gas of bosons with several internal states
leads to collective transport of internal polarization. The internal 
dynamics can be understood as Larmor precession in the presence of 
a torque induced by atoms on each other via 
exchange coupling. A generalized Bloch equation that includes 
interatomic exchange effects as well as orbital motion in the gas  
is derived and used to interpret recent experiment 
by Lewandowski et al. \cite{Rb01} as an excitation of 
a collective wave of internal state polarization. It is shown that
exchange leads to formation of domains in which precession frequencies
are synchronized.
\vskip2mm
\end{abstract}
 
\bigskip
 
\begin{multicols}{2}
\narrowtext
 

Atomic gases in the cold collision regime characterized by
de Broglie wavelength long compared to the range of interparticle 
potential 
represent an interesting quantum many-body system. 
Most surprisingly,
spin waves in a cold spin polarized gas are collective excitations.
This phenomenon was actively studied in the 80's, first
predicted by Bashkin \cite{Bashkin81}
and independently by Lhuillier and Lalo\"{e} \cite{Lhuillier82},
and confirmed by NMR experiments in spin polarized $H\!\downarrow$
by Johnson et al.\cite{Johnson84}, 
in $^3{\rm He}$ by Nacher et al. \cite{Nacher84}, 
and in dilute $^3{\rm He}-^4\!{\rm He}$
mixtures by Gully and Mullin \cite{Gully84}.
A detailed quantitative theory of the observed NMR spectra 
was given by Levy and Ruckenstein \cite{Levy84}. 
Bigelow et al. \cite{Bigelow89} demonstrated that
collective spin waves are preserved 
even in the Knudsen regime.
The theory was further developed by Miyake et al. \cite{Miyake85}
and reviewed in \cite{Bashkin86}.

Exchange effects in gases are not limited to spin phenomena, since 
any pair of internal states can play a role similar to spin states
in exchange collisions \cite{Verhaar95}. 
Apart from new energy scales arising due to internal states spectrum, 
the main difference is in the anisotropic character of exchange. 
For generic internal states 
the Hamiltonian does not have spin-rotational symmetry, which opens
a number of new interesting possibilities. Verhaar et al. \cite{Verhaar87} 
demonstrated that
interatomic exchange 
leads to enhancement (by a factor of two) of the density shift
of Rabi transition. 
Tiesinga et al. \cite{Tiesinga92} and Kokkelmans et al. \cite{Kokkelmans97}
considered application of
this effect in Cesium fountain clock. 
Similar exchange enhancement occurs in the optical 
spectrum density shift. The theory \cite{Bashkin81,Lhuillier82}
of collective spin-waves, can be straightforwardly
adapted to describe optical excitations \cite{Oktel99}.

New aspects of cold collision exchange 
arise in experiments on Bose-Einstein condensation (BEC)
in trapped gases. 
The exchange part of the density shift is absent in BEC at $T=0$ and is
reduced at $0<T<T_{\rm BEC}$ \cite{Oktel99,Oktel00,Pethick01}. Interestingly, 
in this case 
all modes involving coupling of internal states 
are split into doublets \cite{Oktel01}.

In a recent experiment \cite{Rb01} Rabi transition was studied in a
cigar shaped sample of Rb vapor contained in Ioffe-Pritchard trap.
The $|F,\,m_f\rangle =|1,-1\rangle,\ |2,1\rangle$
levels of the hyperfine multiplet 
of Rb split by $\omega_0\approx 6.8\,{\rm GHz\,}$ were
used. 
Almost perfect compensation of the density shift by spatially varying
Zeeman frequency was achieved. 
The transition 
frequency varied along the sample axis by few tens of Hz.
Initial inner state with polarization in the $x-y$ plane was prepared 
by a $\frac{\pi}2$ pulse.
It was observed that the polarization {\it does not remain} in the
$x-y$ plane during free Larmor precession. This was argued to result from
spatial segregation of atoms with different $z$ spin components.
However, the confinement potential 
\cite{Rb01} spin dependence was very weak, and the best estimate 
of segregation due to the difference in mechanical forces experienced
by atoms with different spins was at least an order of magnitude
longer than the time $\approx 0.2\,{\rm s\,}$ of the 
$z$ component buildup.

We argue below that the phenomena of Ref.\cite{Rb01} are explained 
by 
coherent evolution of atoms' internal state rather
than by mechanical segregation
in the gas. The
observed $z$ component profile is readily accounted for by
interatomic exchange coupling.
The transition frequency \cite{Rb01} varies along the
sample axis, and a short time after 
precession started a gradient of precession angle builds up.
Now, consider two interacting atoms with
slightly different polarization due to spatially varying
Larmor frequency.
The exchange interaction of these atoms can be described \cite{Verhaar95}
as precession of each atom's spin around the net spin of both atoms.
Since both atoms have transverse polarization, the precession
about a net spin (which is also transverse) will move
the spins out of the $x-y$ plane and both of them will acquire
a finite $z$ component.

Exchange effects can be illustrated by a thought experiment
involving a gas of identical atoms with density $n$ 
and spin $1/2$ contained in a box. Take the spin polarization 
$\vecs s$ to be purely transverse and the same for all atoms. 
For isotropic exchange coupling 
$
{\cal H}=\hbar \int \lp \omega_0 s^z(r)
+\frac{\lambda}2\vecs s(r)\cdot\vecs s(r)\rp d^3r
$
the polarization $\vecs s$ is uniformly precessing, 
$
\vecs s(t)={\textstyle \frac12} n
\lp \cos\omega_0t\, \hat{\vecs x}+\sin\omega_0t\, \hat{\vecs y}\rp
$. Now
consider a test atom passing through the box with spin
polarization different from that of the other atoms. 
The test atom spin will experience an effective `magnetic' field
$\vecs B= \omega_0\hat{\vecs z} + \lambda  \vecs s(t)$ with 
the exchange part $\lambda  \vecs s(t)$ giving rise to Rabi transitions. 
In a Larmor coordinate frame
rotating with frequency $\omega_0$ about the $z$ axis the effective field is 
just $\lambda n$ along the gas polarization $\vecs s$, 
time-independent in this frame. Since $\vecs s$
is transverse, Rabi transition 
will generate a $z$ component of the test atom polarization, 
even if initially it was in 
the $x-y$ plane. 

Before accepting this explanation 
one needs 
to discuss energy conservation. The probabilities to find the test atom in 
the up and down states after coming out of the box differ from
those in the initial state, since its $z$ spin component 
changes. 
This means that the test atom energy 
can change by $\hbar\omega_0$. 
The total energy of the system, however, does not change because 
the spins 
coupled by exchange precess together around the net spin  
so that the total spin is conserved \cite{Verhaar95}. The change of 
the net spin $z$ component in the box is equal and 
opposite to the test atom spin change, as required by the energy balance. 

Although everything is consistent with energy conservation,
the energy change of $\hbar\omega_0$ with  
$\omega_0\approx 6.8\:{\rm GHz\,}$ much higher than other 
frequencies in the system \cite{Rb01} 
may appear counter-intuitive. The temperature $T= 600\, {\rm nK}$
\cite{Rb01} corresponds to
$k_B T/\hbar \sim 10 \: {\rm KHz}$, the 
trap frequencies are
$(\omega_\perp,\omega_z) = (230,7)\: {\rm Hz}$. However, the characteristic
exchange frequency is 
$\lambda n \simeq 140\: {\rm Hz}$ for the typical density 
$n = 2\: 10^{13} \: {\rm cm}^{-3}$ (see below). This is much higher than  
the transition frequency broadening estimated from the precession decay time 
to be of order of few Hz \cite{Rb01}. 
This makes the exchange induced Rabi transitions at the energy 
$\hbar\omega_0$ fully coherent, despite that $\hbar\omega_0\gg \lambda n$. 

The length corresponding to one Rabi cycle is
%
\be
l_{\rm exch} = \frac{v_{\rm T}}{\lambda n} = 16 \, {\rm \mu m}
,\quad \lp v_{\rm T}=\sqrt{2T/m}\rp
\ee
This is larger than the sample radius
$r_\perp = 7.3 \, {\rm \mu m}$ but
much smaller than the sample length
$r_z = 240\, {\rm \mu m}$. Since movement of an atom 
by $\simeq \frac14 l_{\rm exch}$ is sufficient for rotating the spin 
by $\frac{\pi}2$ and moving it out of the $x-y$ plane, 
the exchange coupling is a viable mechanism 
for spin reorientation in this system.

The separation of Rb atoms into a gas sample and a test
particle in the thought experiment is artificial.
The atoms in \cite{Rb01} share both roles, by 
inducing precession on each other via exchange coupling. One therefore has 
to consider {\it collective dynamics} of atom polarization 
\cite{Bashkin81,Lhuillier82}. 
The Hamiltonian of Rb atoms in a trap has the form
\bea
&&{\cal H}=\int\lp\sum_{j=1,2} \bar\psi_j {\it H}_j\psi_j
+ \frac{\hbar}2 \sum_{j,k=1,2} \lambda_{jk} :\!\hat n_j\hat n_k\!: \rp d^3r
\\
&&{\it H}_j =  -{\textstyle \frac{\hbar^2}{2m}}\nabla^2
+U_j( r)\ ,\quad \lambda_{jk}={\textstyle \frac{4\pi\hbar}{m}} a_{jk}
\eea
where $\hat n_j=\bar\psi_j\psi_j$ is 
the density operator. For the states used 
in Ref.\cite{Rb01} the scattering lengths are
$(a_{11},a_{22},a_{12}) = (100.9, 95.6, 98.2) \: a_0 $
with $a_0$ the Bohr's radius.  

The polarization of internal states is described by `spin' operators
with components given by Pauli matrices
\be\label{eq:Sxyz}
\hat{s}^{x(y,z)}(r)=
\frac12\sum_{j,k}\bar\psi_j(r)\sigma^{x(y,z)}_{jk}\psi_k(r) 
\ee
and standard commutation algebra 
\be\label{eq:Salgebra}
\lb \hat{s}^{\alpha}(r),\hat{s}^{\beta}(r')\rb 
= i\,\varepsilon_{\alpha\beta\gamma}\hat{s}^{\gamma}(r)\delta(r-r')
\ee
of spin density operators.

The system \cite{Rb01} is deep in the cold collision regime, 
since thermal de Broglie wavelength $\lambda_{\rm T}=h/mv_{\rm T}
\simeq 4000 \: a_0$ is much larger than the scattering lengths
$a_{jk}$. We employ the forward scattering approximation
also known as the random phase approximation \cite{RPA}. 
The interaction can be rewritten
in momentum representation as
%
\bea\label{eq:RPA}
&&\int :\!\hat n_j\hat n_k\!: d^3r \ = \sum_{p+p'=p''+p'''} 
\bar\psi_{j,p}\bar\psi_{k,p'}\psi_{j,p''}\psi_{k,p'''} 
\\ \nonumber
&&
=\! \sum_{p,p'\!,q}
\lp\! \bar\psi_{j,p_+}\!\bar\psi_{k,p'_-}\!\psi_{j,-p_-}\!\psi_{k,-p'_+} 
\!+\! \bar\psi_{j,p_+}\!\bar\psi_{k,p'_-}\!\psi_{j,-p'_+}\!\psi_{k,-p_-}\! \rp
\eea
%
\noindent
where $p_\pm=p\pm q/2$. 
The first term of (\ref{eq:RPA}) accounts for the forward scattering process,
while the second term describes exchange scattering. Identifying the 
operators in (\ref{eq:RPA}) with the spin density components
(\ref{eq:Sxyz}) we obtain
\bea
:\!\hat n_1\hat n_1\!: \ ={\textstyle \frac12} (\hat n + 2\hat s^z)^2
,\quad
:\!\hat n_2\hat n_2\!: \ ={\textstyle \frac12} (\hat n - 2\hat s^z)^2
\\
:\!\hat n_1\hat n_2 + \hat n_2\hat n_1\!:
\ ={\textstyle \frac12}  \hat n^2  - 2(\hat s^z)^2
+ 4\hat s^+ \hat s^- + 4\hat s^- \hat s^+
\eea
%
In the spin representation the interaction
has the form
\be
{\textstyle \frac12}\sum_{j,k} \lambda_{jk} :\!\hat n_j\hat n_k\!:
= {\textstyle \frac{u}2}\hat n^2
+\Lambda\, \hat n  \hat s^z 
+ \delta\lambda\, (\hat s^z)^2 + \lambda_{12} \vecs s^2
\ee
with $u=\lambda_{11}+\lambda_{22}+\lambda_{12}$, 
$\Lambda=\lambda_{11}-\lambda_{22}$, and 
$\delta\lambda=\lambda_{11}+\lambda_{22}-2\lambda_{12}$. 
Spin dynamics is given by
$\partial_t {\hat{\vecs s}}= \frac{i}{\hbar} \lb {\hat{\vecs s}},
{\cal H}\rb $, where
the commutator can be evaluated with the help of the relations 
(\ref{eq:Salgebra}). After taking the expectation values $\vecs s = \langle \hat{\vecs s}\rangle$ we obtain
a generalized Bloch equation
\be\label{eq:BlochS}
\partial_t \vecs s + \vec \nabla \cdot \vec{\vecs j} = \vecs \Omega \times \vecs s
,\quad 
\vecs \Omega = (\omega_0 +\delta\omega) \hat{\vecs z} + 2\lambda_{12}\vecs s
\ee
with $\vec{\vecs j}(r)=
-\frac{i\hbar}{2m}\langle \bar\psi_j\vecs s_{jk}\vec\nabla \psi_k\rangle 
+ {\rm h.c.}$ the spin current. Here
\be \label{eq:Omegaz}
\delta\omega(r)={\textstyle \frac1{\hbar}}(U_1-U_2)
+\Lambda\, n
 + 2\,\delta\lambda\, s^z
\ee
To make contact with the discussion in Ref.\cite{Rb01} we note that 
$\frac12 n\pm s^z=n_{1(2)}$, the occupation probabilities for 
the up and down spin. Combined with the form of $\Lambda$ and $\delta\lambda$,
the frequency $\delta\omega(r)$ 
can be rewritten
as
\be
\delta\omega(r)\! =\! {\textstyle\frac1{\hbar}}(U_1\!-\!U_2)
\!+\! 2(\lambda_{11}\! -\! \lambda_{12})n_1
\!-\! 2(\lambda_{22}\! -\! \lambda_{12})n_2 
\ee
The first term is the Zeeman frequency shift due to 
the trap field inhomogeneity, while the last two terms 
(identical to Eq.(1) of Ref.\cite{Rb01}) give the density shift.
The term $2\lambda_{12}\vecs s$ in the expression
(\ref{eq:BlochS}) for $\vecs \Omega(r)$ representing 
the effect of exchange
is not considered in Ref.\cite{Rb01}. The role of this term is subtle. 
It drops out from the 
Bloch equation (\ref{eq:BlochS}) for $\vecs s$, 
since $\vecs s \times \vecs s=0$. However, since typically
$2\lambda_{12}|\vecs s|\gg |\delta\omega(r)|$, 
this term should be taken into account in the Bloch equation for 
other spin-related quantities, such as the spin current $\vec{\vecs j}$. 
The torque $\vecs s\times \vecs j$ (which in general is
not along the $z$ axis) makes the spin current precess so that
the $x$, $y$ and $z$ components intermix.

The transport equation for the spin current $\vec{\vecs j}$ is derived 
in a similar fashion \cite{Bashkin81,Lhuillier82}. We obtain
\be\label{eq:J}
\partial_t \vec{\vecs j} + \alpha \vec\nabla \vecs s 
= \vecs \Omega(r) \times \vec{\vecs j} - \gamma\, \vec{\vecs j}
,\quad
\alpha={\textstyle \frac13} v_{\rm T}^2
\ee
where the elastic collision rate $\gamma = 4\pi a^2 v_{\rm T} n\simeq 20\,{\rm Hz}$ 
with parameters of Ref.\cite{Rb01}.
The term $\alpha \nabla \vecs s$ arises
in a standard way after retaining angular harmonics of the lowest
order in the transport equation. In 
Eq.(\ref{eq:J}) we ignored the terms such as
$\vecs s \nabla (U_1+U_2)$ and $n \hat{\vecs z} \nabla (U_1-U_2)$, 
since their magnitude is small
(see Ref.\cite{Rb01}). In this approximation, the spin 
and density dynamics decouple, in agreement with 
the observation \cite{Rb01}.

Since the interaction parameters $\lambda_{jk}$ for Rb coincide within $3\%$,
spin is approximately conserved 
by elastic collisions. In this case, the spin current relaxation 
rate $\gamma$ is the same as for the average particle momentum, and 
no spin relaxation appears in the Bloch equation 
(\ref{eq:BlochS}). Elastic collisions control spin 
relaxation indirectly, by making spin diffusion constant complex
\cite{Bashkin81,Lhuillier82}.

The transport equations (\ref{eq:BlochS}),(\ref{eq:J}) 
can be simplified for a one dimensional
system \cite{Rb01} by averaging over sample crossection. 
Large  
exchange $\lambda n \simeq \omega_\perp$ \cite{Rb01} 
leads to fast dynamical averaging of spin polarization 
in each crossection with parameters slowly varying along the sample length. 
In averaging 
Eqs.(\ref{eq:BlochS}),(\ref{eq:J}) 
we assume Gaussian density profile 
$n(\rho)=n e^{-\rho^2/r_\perp^2}$. The averaging of
the terms in Eqs.(\ref{eq:BlochS}),(\ref{eq:J}) 
quadratic in density and/or spin is performed as 
$\int n^2(\rho) d^2\rho/\int n(\rho) d^2\rho = \frac12 n$, 
where $n$ is the peak density.
After rescaling
all coupling constants 
\be
\lambda_{jk}\to {\textstyle\frac12}\lambda_{jk}
\ee
and replacing $\nabla$ by one dimensional $\partial_x$ we obtain 
transport equations of Leggett-Rice form \cite{Leggett68}
%
\bea \label{eq:S1D}
&& \partial_t \vecs s + \partial_x\vecs j = (\omega_0 + \widetilde{\delta\omega}(x))\, \hat{\vecs z} \times \vecs s
\\ \label{eq:J1D}
&& \partial_t \vecs j + a \partial_x\vecs s 
= \lp (\omega_0 + \widetilde{\delta\omega}(x))\, \hat{\vecs z} + \lambda_{12}\vecs s \rp \times \vecs j - \gamma\, \vecs j
\\ \label{eq:tildeOmega}
&& 
\widetilde{\delta\omega}(x)
={\textstyle \frac1{\hbar}}(U_1-U_2)
+\frac12\Lambda n
 + \delta\lambda\,s^z
\eea
The coupled dynamics of $\vecs s$ and $\vecs j$ is nonlinear because 
of the exchange precession 
torque $\lambda_{12}\vecs s \times \vecs j$ in Eq.(\ref{eq:J1D}). 

In the approximation 
$\widetilde{\delta\omega}(x), \gamma \ll \lambda_{12}n$ \cite{Rb01} one can
simplify transport equations by performing 
a gradient expansion. We first go to the Larmor frame rotating with 
frequency $\omega_0$, which eliminates $\omega_0$ from 
Eqs.(\ref{eq:S1D}),(\ref{eq:J1D}). Next, ignoring the time derivative 
$\partial_t \vecs j$ in Eq.(\ref{eq:J1D}) we solve it 
for $\vecs j$ in terms of $\vecs s$ and $\partial_x\vecs s$, and substitute 
the result in Eq.(\ref{eq:S1D}).
This gives the Landau-Lifshitz equation \cite{Landau35}
\bea\label{eq:L-L}
{\displaystyle \partial_t \vecs s - 
\partial_x\lp D_1(\vecs s) \partial_x\vecs s \rp 
= [ \widetilde{\delta\omega}(x)\,\hat{\vecs z} - D_2(\vecs s)\partial_x^2\vecs s ] \times \vecs s
}
\\
D_1(\vecs s)= \frac{\alpha \gamma}{\gamma^2+\lambda_{12}^2 \vecs s^2}
,\qquad 
D_2(\vecs s)= \frac{\alpha \lambda_{12}}{\gamma^2+\lambda_{12}^2 \vecs s^2}
\eea
It is convenient to nondimensionalize Eq.(\ref{eq:L-L}). 
We rescale $\vecs s$ by $|\vecs s|_{\rm max}=\frac12 n$, the frequency 
$\widetilde{\delta\omega}(x)$ and scattering rate $\gamma$ by 
$\lambda_{12}|\vecs s|_{\rm max}$, and choose as length unit
\be\label{eq:Lunit}
\frac{\lambda_{12}|\vecs s|_{\rm max}}{\alpha^{1/2}}
\equiv \frac{\lambda_{12} n}2 \lp\frac{3m}{2k_{\rm B}T}\rp^{1/2}
\equiv \frac{\sqrt{3}}2 l_{\rm exch}
\approx 14\,{\rm \mu m}
\ee
Eq.(\ref{eq:L-L}) 
preserves its form, with $D_2=1/(\gamma^2+\vecs s^2)$
and $D_1=\gamma D_2$. The dimensionless damping $\gamma$ is 
\be
\gamma = \frac{4\pi a^2 v_{\rm T}n}{\frac12\lambda_{12}n}
=\frac2{\hbar} a m v_{\rm T}
=4\pi\frac{a}{\lambda_{\rm T}}
\simeq 0.31
\ee
since $a\approx 100\,a_0$ and the de Broglie wavelength
$\lambda_{\rm T}=h/mv_{\rm T}\approx 4000\,a_0$ \cite{Rb01}.

\begin{figure}
\centerline{\psfig{file=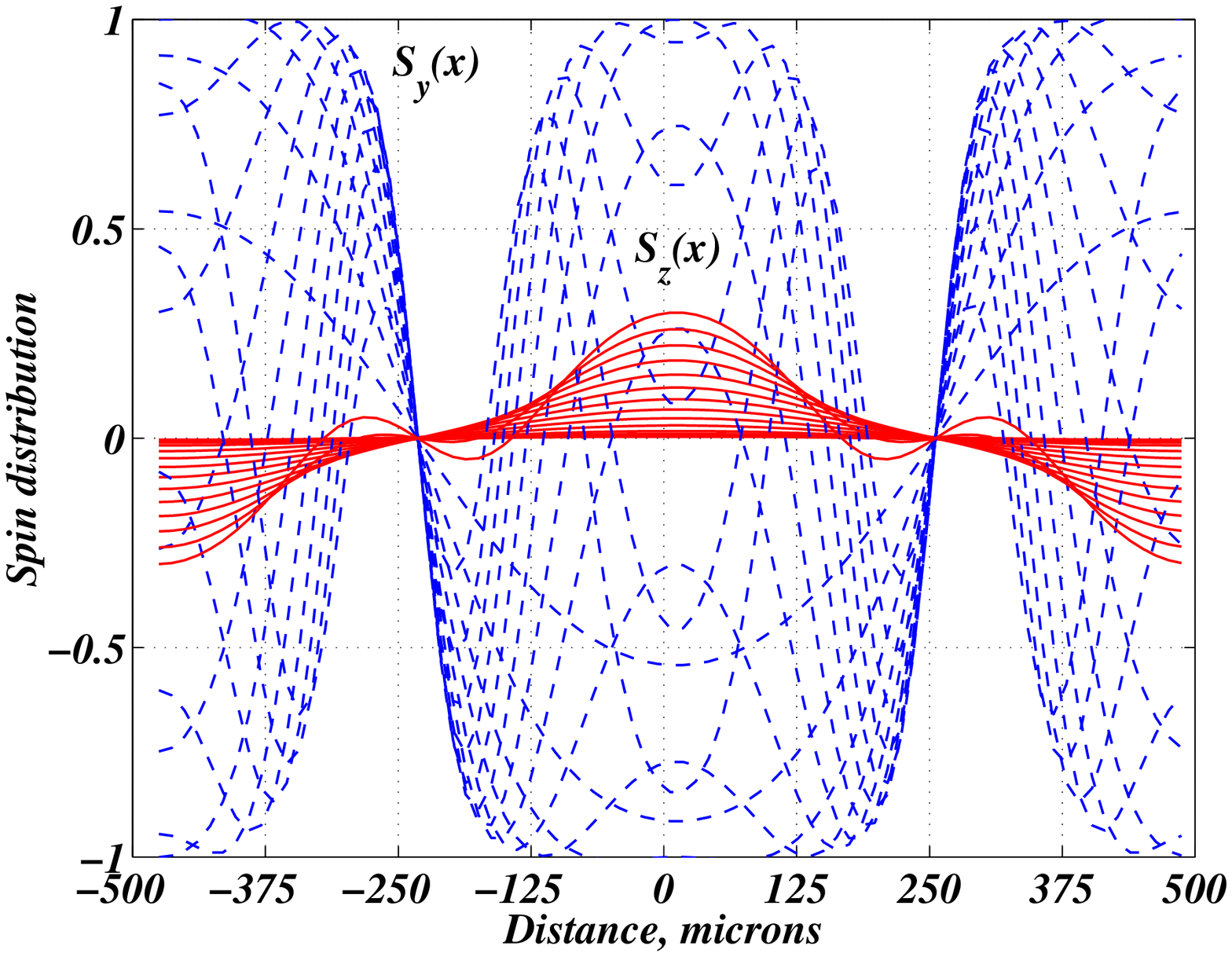,width=3in,height=2in}}
\vspace{0.25cm}
\centerline{\psfig{file=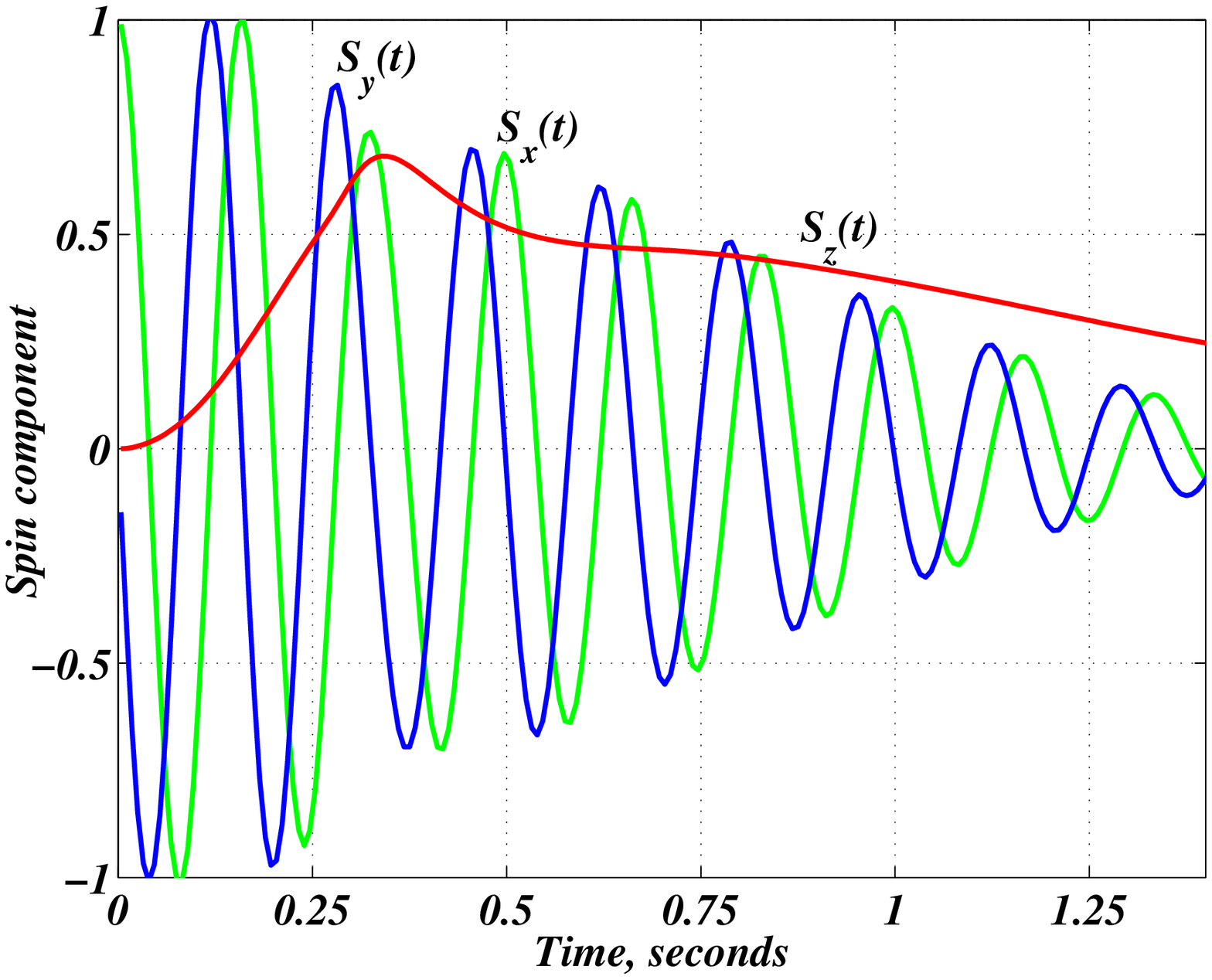,width=3in,height=2in}}
\vspace{0.25cm}
         \caption[]{
Numerical simulation of Eq.(\ref{eq:L-L}). Parameters used:
constant density $n=2\,10^{13}\,{\rm cm}^{-3}$, 
exchange frequency $\lambda_{12}|\vecs s|_{\rm max}= 70\,{\rm Hz}$.
Spatial variation of the transition frequency 
$\widetilde{\delta\omega}(x)=-\Omega \cos(2\pi x/L)$
with $\Omega=20/\pi\approx 6.37\,{\rm Hz}$ and sample size $L=10^3\,{\rm \mu m}$. 
Top: spin distribution  
evolution at $0<t<0.2\,{\rm sec}$; 
Bottom: time dependence at $x=0$.
        }
\label{fig:L-L}
\end{figure}

The results of numerical simulation of Eq.(\ref{eq:L-L}) are shown 
in Fig.\ref{fig:L-L}. The spatial and temporal behavior 
is similar to that in Ref.\cite{Rb01}: The $z$ component builds up
$\simeq 0.2\,{\rm s}$ after precession started and 
then gradually decays to zero along with the oscillating transverse 
component. 

\begin{figure}
\centerline{\psfig{file=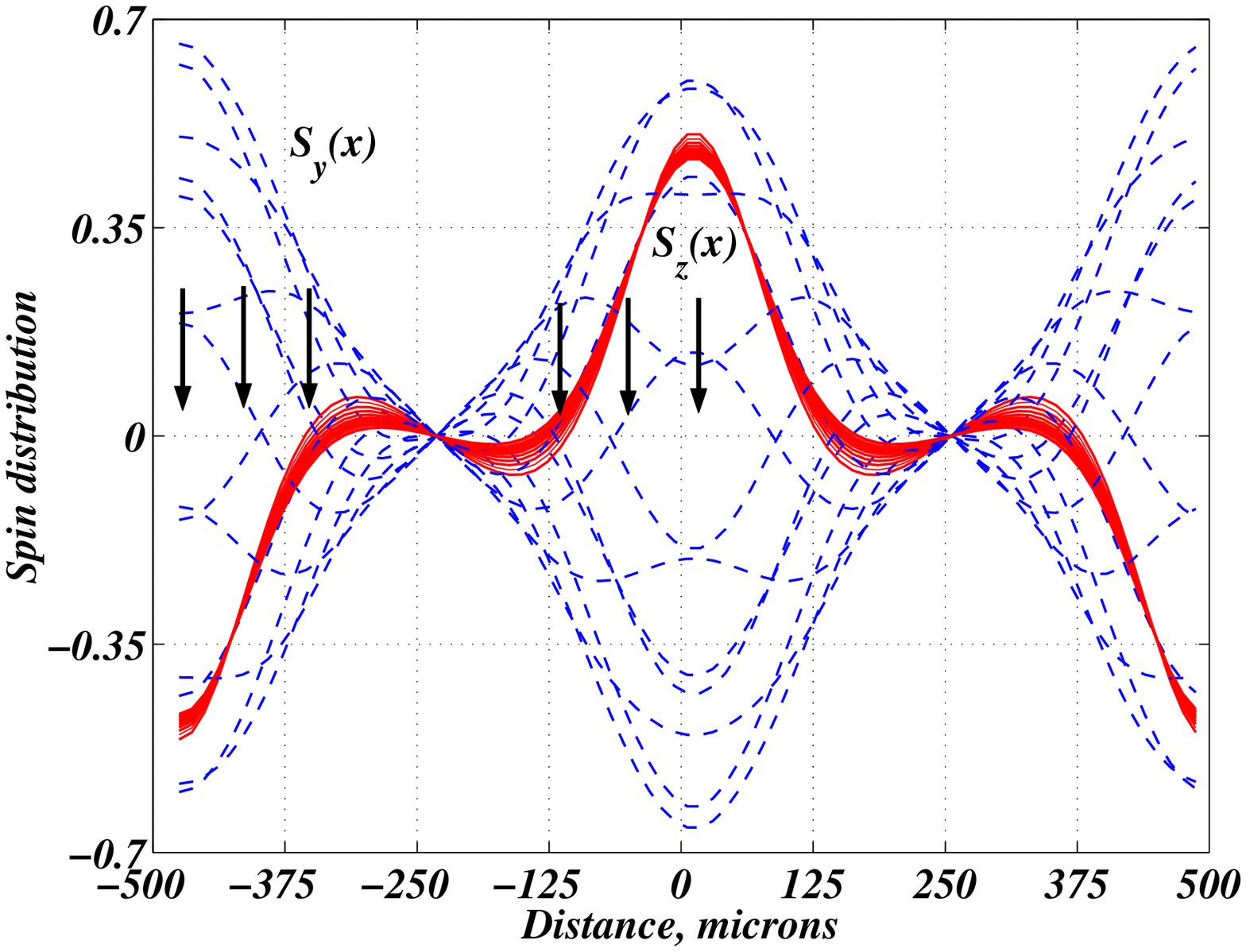,width=3in,height=2in}}
\vspace{0.25cm}
\centerline{\psfig{file=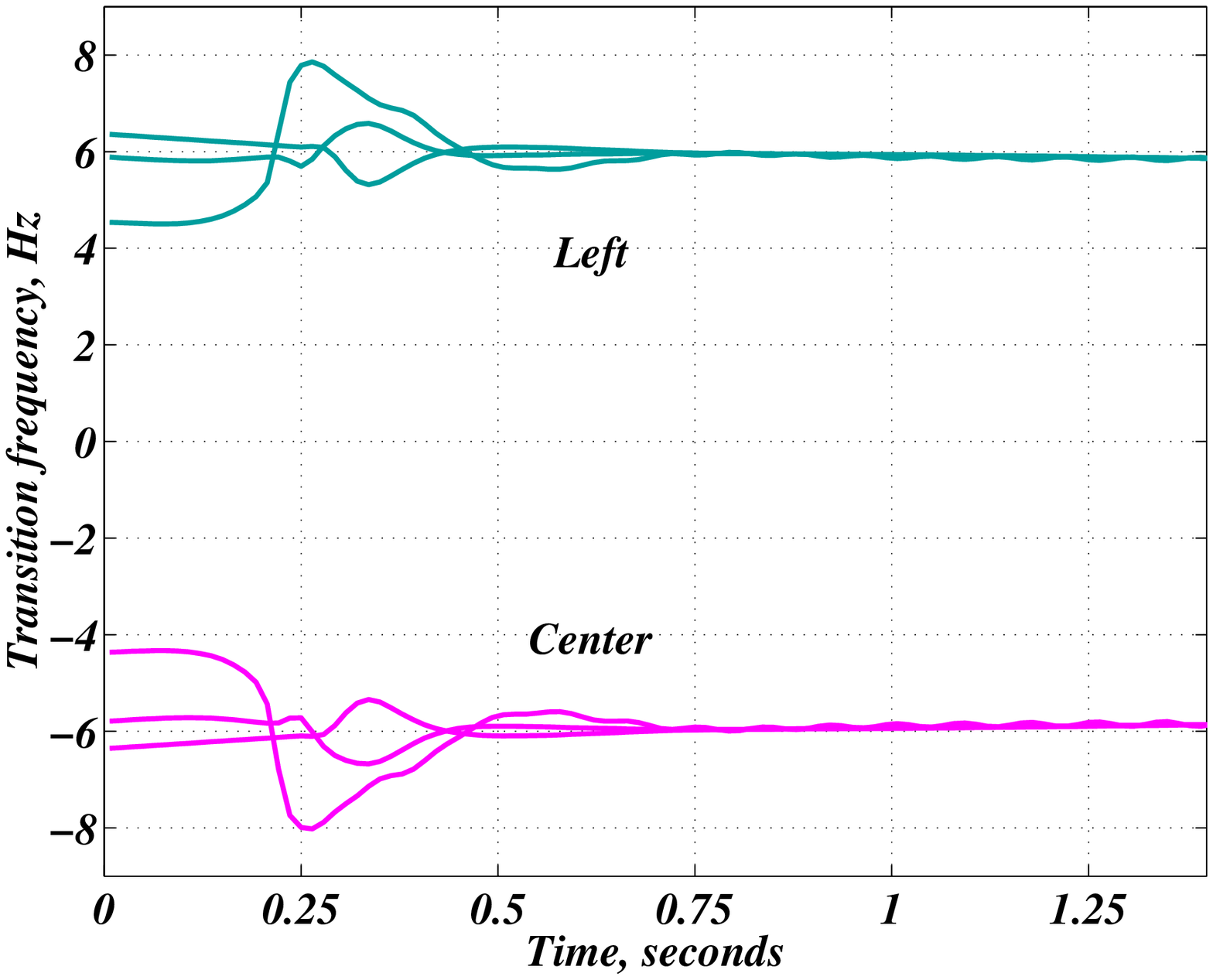,width=3in,height=2in}}
\vspace{0.25cm}
         \caption[]{
Top: spin distribution evolution from $0.5\,{\rm sec}$ to $0.7\,{\rm sec}$
(with the same parameters as in Fig.\ref{fig:L-L});
Bottom: frequency 
(\ref{eq:freqXY}) 
synchronization 
at the points marked by arrows.
        }
\label{fig:synchro}
\end{figure}

Spin precession
becomes {\it synchronized} 
in different parts of the sample (see Fig.\ref{fig:synchro}), 
due to compensation of the transition frequency 
$\widetilde{\delta\omega}(x)$ spatial variation 
by the exchange field $\lambda_{12}\vecs s(x)$. In our simulation, 
synchronization takes place independently 
in the domains with $s_z>0$ and $s_z<0$. 
Frequency was evaluated as
\be\label{eq:freqXY}
f={\textstyle \frac1{2\pi}}d\theta/dt,\qquad  \theta=\,{\rm arg\,}(s_x+is_y)
\ee
During the first $0.2\,{\rm s}$ of the $z$ component buildup
the frequency evolves from initial value 
$f={\textstyle \frac1{2\pi}}\widetilde{\delta\omega}(x)$ 
to a constant value $\approx \pm 6\,{\rm Hz}$ in each domain.

While precession frequencies become synchronized,
the phase $\theta$ varies within each domain producing spin flux
between the domains. Spin density $\vecs s(x)$ vanishes at the 
domain boundaries $x=\pm\frac14L=\pm 250\,{\rm \mu m}$ 
(see Figs.\ref{fig:L-L},\ref{fig:synchro}).
The number of synchronized domains
and domain-specific frequency values  
in general depend on the amplitude and characteristic spatial 
scale of $\widetilde{\delta\omega}(x)$. 

The mechanism of transverse spin component decay in the synchronized 
state is polarization mixing caused by spin current between 
different domains.
The time scale of spin decay, set by spin diffusion,
is much longer than the elastic collision time. The $z$ component 
first builds up due to spin currents and then decays due to 
(longitudinal) diffusion, with characteristic time $(L/2\pi)^2\gamma/\alpha
\approx 0.5\,{\rm secs}$ (see Fig.\ref{fig:L-L}). This is consistent with Ref.\cite{Rb01}. 

In summary, exchange coupling in a trapped gas leads to complex collective 
dynamics of polarization. Polarized atoms exert torque on the spin 
current creating a $z$ component 
profile in the presence of spatially varying transition 
frequency. Results of numerical simulation of the dynamics 
of the Rb system \cite{Rb01} are in agreement with observations. 
Surprisingly, the buildup of the $z$ component
is accompanied by synchronization of precession frequencies. 
In the inhomogeneous state the sample breaks into two or more 
synchronized domains.
Spin relaxation is caused by spin currents between the domains. 

Synchronized precession should manifest itself 
in experiment as transition frequency locking to one value 
in the entire sample, if it is a single domain. Several synchronized 
domains formed within the sample will give rise to several plateaus 
in the transition frequency spatial dependence. 
Spin density vanishing between different domains should be observable 
by the spatially resolved Ramsey fringes technique of Ref.\cite{Rb01}

We are grateful to E. A. Cornell and Tin-Lun Ho for useful discussions.

\end{multicols}
\end{document}